\documentclass{article}

\usepackage{arxiv}

\usepackage{amsmath,amssymb,amsfonts}
\usepackage{algorithm}
\usepackage{algorithmicx}
\usepackage{algpseudocode}
\usepackage{cite}
\usepackage{epsfig}
\usepackage{float}
\usepackage{graphicx}
\usepackage{hyperref}
\usepackage{mathtools}
\usepackage{natbib}
\usepackage{textcomp}
\usepackage{pgf}
\usepackage{subcaption}
\usepackage{siunitx}
\usepackage{tikz}
\usepackage{xcolor}
\usetikzlibrary{shapes.geometric, arrows, positioning}

\usepackage{enumitem}
\setlist[itemize]{left=0pt,labelsep=5pt}

\newcommand{\Title}{Reinforcement Learning-Based Market Making as a Stochastic Control on Non-Stationary Limit Order Book Dynamics}

\title{\Title}

\date{}

\author{%
  \href{https://orcid.org/0009-0008-6064-9895}{\includegraphics[scale=0.06]{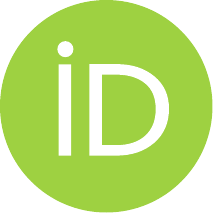}\hspace{1mm}Rafael Zimmer} \\
  Institute of Mathematics and Computer Sciences\\
  University of São Paulo\\
  \texttt{rafael.zimmer@usp.br}
  \And
  \href{https://orcid.org/0000-0001-5989-7287}{\includegraphics[scale=0.06]{orcid}\hspace{1mm}Oswaldo L. V.~Costa} \\
  School of Polytechnic Engineering\\
  University of São Paulo\\
  \texttt{oswaldo.costa@usp.br}
}%

\renewcommand{\undertitle}{}
\renewcommand{\shorttitle}{}
\renewcommand{\headeright}{}

\hypersetup{
pdftitle={Reinforcement Learning-Based Market Making as a Stochastic Control on Non-Stationary Limit Order Book Dynamics},
pdfsubject={q-fin.TR, cs.AI},  
pdfauthor={Rafael Zimmer, Oswaldo L. V.~Costa},
pdfkeywords={Reinforcement Learning, Market Making, Limit Order Book, Stochastic Control, Market Microstructure},
}

\begin{document}
    \twocolumn[
    \maketitle
    \vspace{-3em}
    \begin{abstract}
        Reinforcement Learning has emerged as a promising framework for developing adaptive and data-driven strategies,
        enabling market makers to optimize decision-making policies based on interactions with the limit order book environment.
        This paper explores the integration of a reinforcement learning agent in a market-making context,
        where the underlying market dynamics have been explicitly modeled to capture observed stylized facts of real markets,
        including clustered order arrival times, non-stationary spreads and return drifts, stochastic order quantities and price volatility.
        These mechanisms aim to enhance stability of the resulting control agent,
        and serve to incorporate domain-specific knowledge into the agent policy learning process.
        Our contributions include a practical implementation of a market making agent based on the Proximal-Policy Optimization (PPO) algorithm,
        alongside a comparative evaluation of the agent's performance under varying market conditions via a simulator-based environment.
        As evidenced by our analysis of the financial return and risk metrics when compared to a closed-form optimal solution,
        our results suggest that the reinforcement learning agent can effectively be used under non-stationary market conditions,
        and that the proposed simulator-based environment can serve as a valuable tool for training and
        pre-training reinforcement learning agents in market-making scenarios.
    \end{abstract}
    \keywords{Reinforcement Learning \and Market Making \and Limit Order Book \and Stochastic Control \and Market Microstructure}
    \vspace{2em}
    ]

    \section{Introduction}
\label{sec:introduction}

Market making in financial markets can be defined as continuously quoting bid and ask prices to profit from thespread,
and is an essential part of the market microstructure, as it helps to narrow bid-ask spreads, reduce volatility, and maintain market stability~\cite{Glosten1985, OHara1995}.
This is particularly important in times of high uncertainty, as market makers provide liquidity, which helps to maintain a fair and efficient market price.
With the advent of electronic trading, placing optimal bid and ask quotes as a market maker is becoming an almost completely automatized task,
but such a transition requires dealing with additional caveats, including slippage, market impact, adverse market regimes~\cite{Cont2010, Bouchaud2018}
and non-stationary conditions~\cite{Gasperov2021}.

The Reinforcement Learning (RL) paradigm has shown promising results for optimizing market making strategies,
where agents learn to adapt their policies through trial and error given a numerical outcome reward score\footnote{Not to be confused with financial returns.},
which is used to evaluate the agent's performance in the environment~\cite{Sutton2018}.
The RL approach is based on the Bellman equation for state values or state-action pair values,
which recursively define the state-value of a policy as the expected return of discounted future rewards.
State-of-the-art RL algorithms, such as Proximal Policy Optimization (PPO) and Soft Actor-Critic (SAC),
solve the Bellman equation by approximating the optimal policy using neural networks to learn either the policy, the value function~\cite{Sutton2018},
or both~\cite{Schulman2015, Mnih2015}, thus enabling the agent to learn the optimal actions according to observed market states~\cite{He2023, Bakshaev2020}.

Using historical data to train RL agents is a common practice in the market making literature, but has some limitations
due to being computationally expensive and requiring large amounts of data,
besides not including the effects of market impact and inventory risk~\cite{Frey2023, Ganesh2019} on the agent's decision-making policies.
An additional approach to creating realistic environments are agent-based simulations,
where generative agents are first trained against observed market messages and then used to simulate order arrivals.
This approach has the disadvantage of trading off fine-grained control over market dynamics for realism,
as well as limiting the agent's adaptability to unseen market regimes, which can lead to underfitting
and suboptimal decision-making under scenarios where market impact and slippage have a more prominent effect on the agent's performance.
On the other hand, using stochastic models to simulate the limit order book environment is computationally cheaper and faster to train,
but may not capture the full complexity of real market dynamics, due to simplified assumptions and static ~\cite{Sun2022}.

In the context of this paper, we propose and discuss a methodology for integrating a reinforcement learning agent in a market-making context,
where the underlying market dynamics are explicitly simulated using parameterizable stochastic processes combined carefully to capture observed stylized facts of real markets.
Our main contribution is implementing multiple non-stationary dynamics into a single limit order book simulator,
and how using fine-controlled non-stationary environments can enhance the resulting agent's performance under adverse market conditions
and provide a more realistic training environment for RL agents in market-making scenarios, while still considering market impact and inventory risk.
We then perform a comparative analysis of the agent's performance under changing market conditions during the
trading day and benchmark it against the Avellaneda-Stoikov market making strategy~\cite{Avellaneda2008} optimal solution (under a simplified market model).

    \section{Bibliography Review}
\label{sec:bibliography-review}

A bibliographical search was performed to determine the current state-of-the-art in market-making research using reinforcement learning.
The analysis performed showed a gap in the literature regarding simulations with varying market regimes and non-stationary environments,
which we aim to address in our research.
We determined the best performing algorithms and state-action spaces used in the literature,
and the most common data types and reward structures used in the references.
The search was made using the Web of Science repository, covering the last 5 years, with the following search key:
\small
\begin{verbatim}
(
    "reinforcement learning" OR
    "optimal control" OR
    "control theory" OR "machine learning"
)
AND
(
    "market making" OR "market maker"
)
\end{verbatim}

Initially, 59 references were selected and deemed relevant to the project,
with 23 of them being effectively used in our analysis, and 5 additional references being added to the list after the initial selection.

\subsection{Data Type}
\label{subsec:data-type}
The majority of the references used historical data, while the second most common type was agent-based simulations,
where generative agents are trained against observed market messages and used to simulate the LOB environment~\cite{Frey2023, Ganesh2019}
trading off control over the market dynamics for replicating observed market flow.

\begin{figure}
    \centering
    \includegraphics[width=0.9\columnwidth]{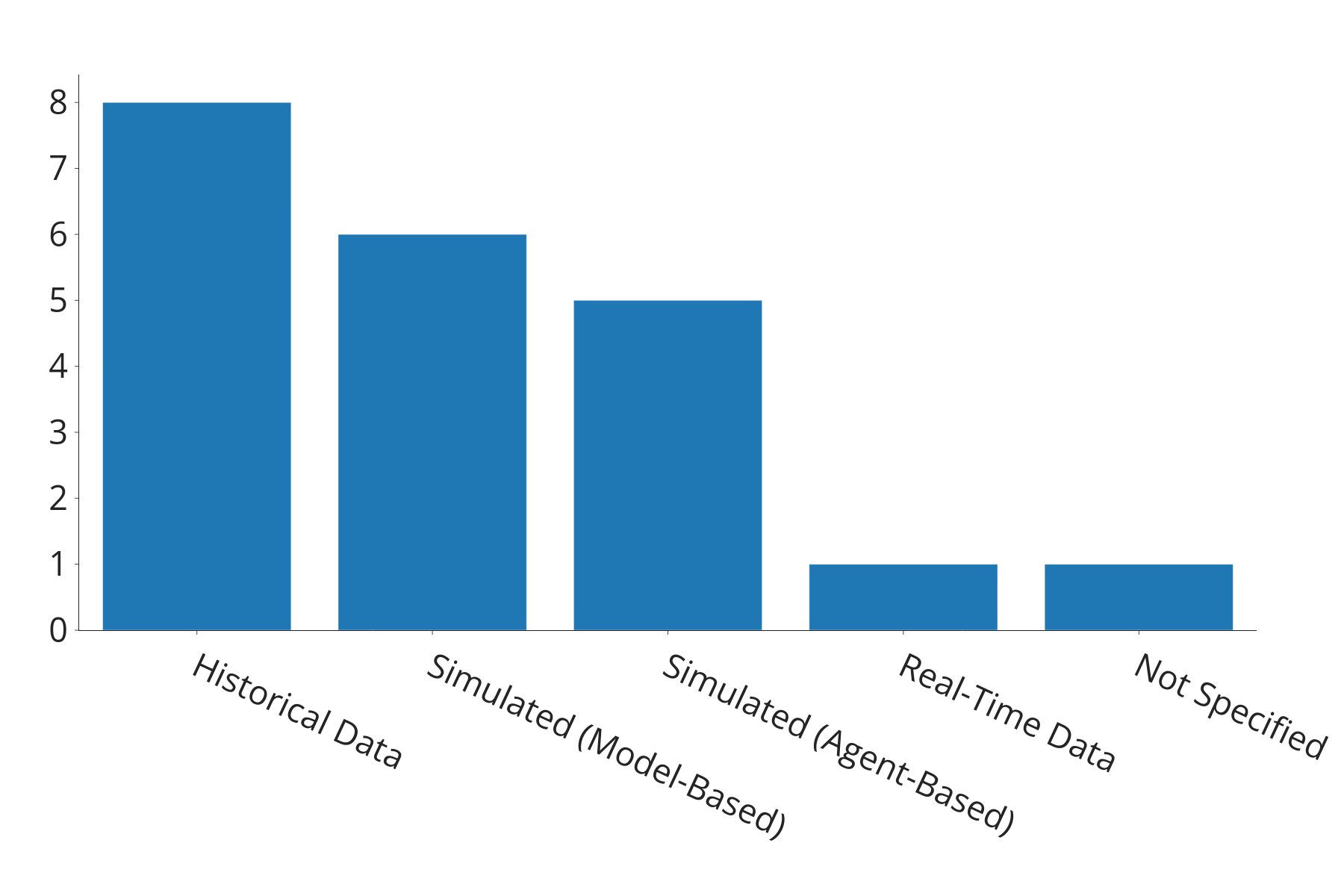}
    \caption{Data types used in the references.}
    \label{fig:figure7}
\end{figure}

The third most common were stationary processes including Geometric Brownian Motion and Poisson processes~\cite{Gasperov2021, Sun2022} for
order arrival and price dynamics, while some used the Hawkes process for simulating order arrival times.
Overall, the references shown in \autoref{fig:figure7} demonstrate a preference for historical data and model-based simulators,
and although being extremely promising approaches, neither completely account for changing market regimes and scenarios where impact and
slippage have a more prominent effect on the agent's performance.
With no references using both simulated data and reinforcement learning agents instead of closed-form solutions,
we aim to address this gap in the literature by implementing a reinforcement learning agent and testing its robustness and generalization capabilities under non-stationary environments.

\subsection{State, Action, and Reward Spaces}
\label{subsec:spaces}

\begin{figure}
    \centering
    \begin{subfigure}{.9\columnwidth}
        \centering
        \includegraphics[width=\linewidth]{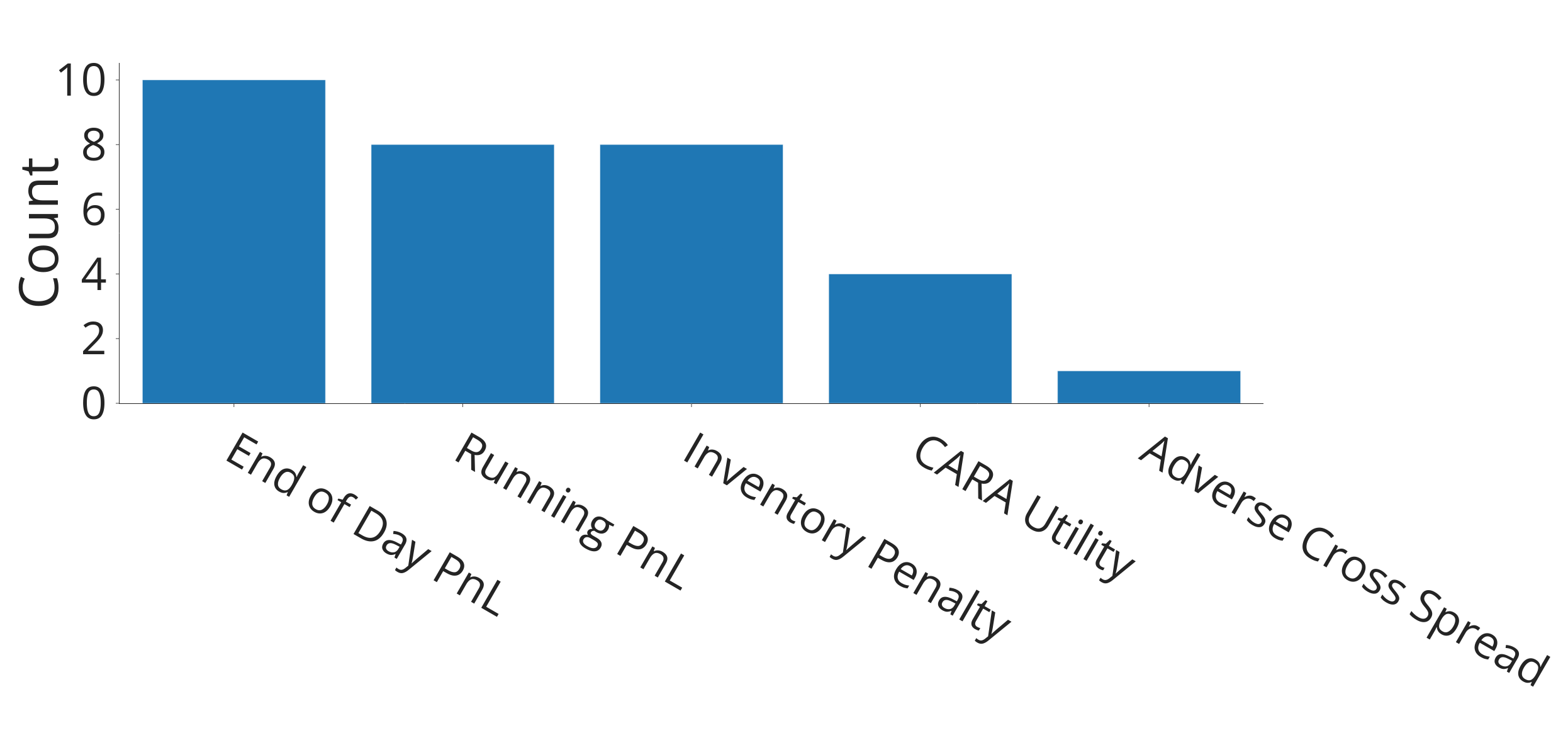}
        \caption{Tagged reward space function variables}
        \label{fig:reward}
    \end{subfigure}
    \vspace{0.5em}
    \begin{subfigure}{.9\columnwidth}
        \centering
        \includegraphics[width=\linewidth]{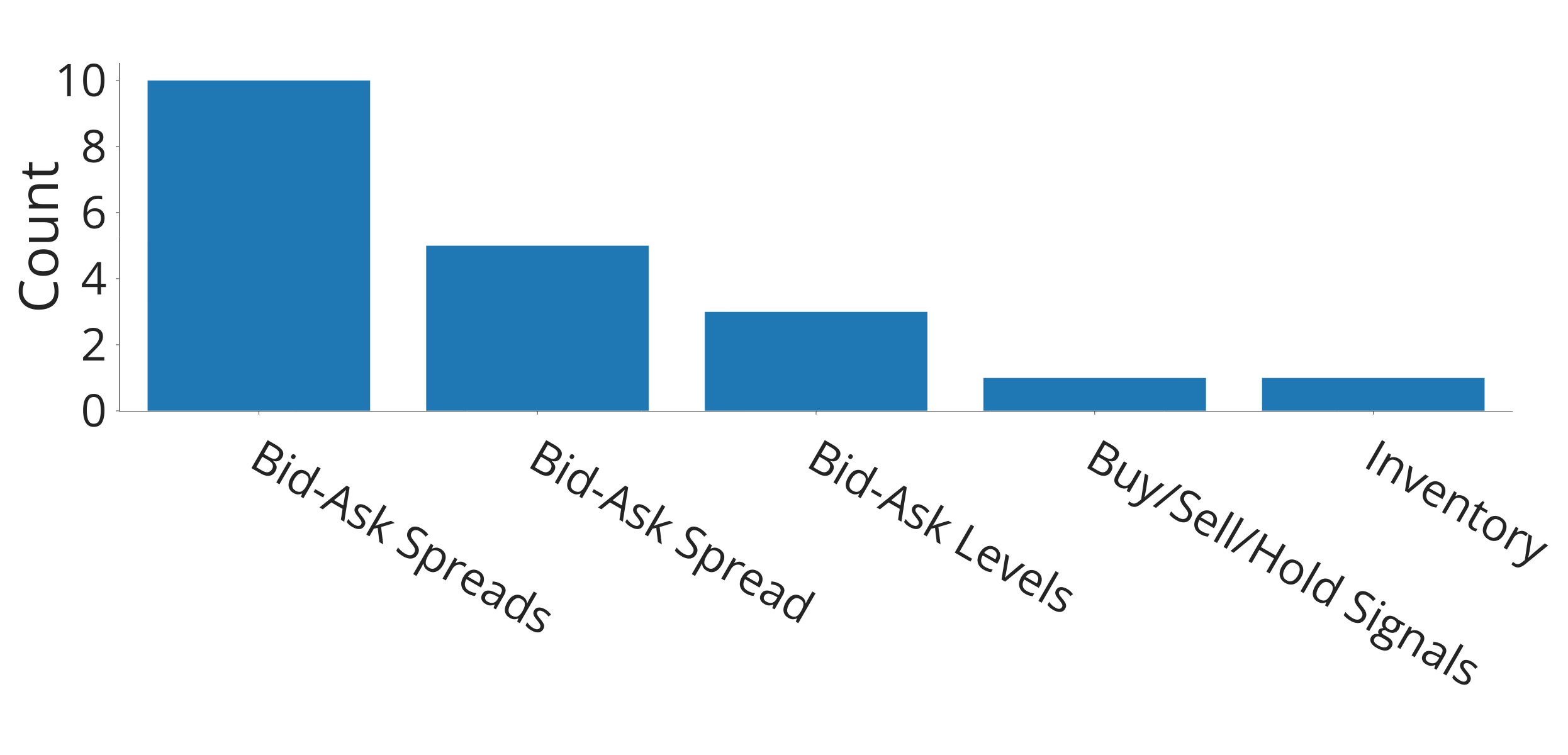}
        \caption{Tagged action space variables}
        \label{fig:action}
    \end{subfigure}
    \vspace{0.5em} 
    \begin{subfigure}{.9\columnwidth}
        \centering
        \includegraphics[width=\linewidth]{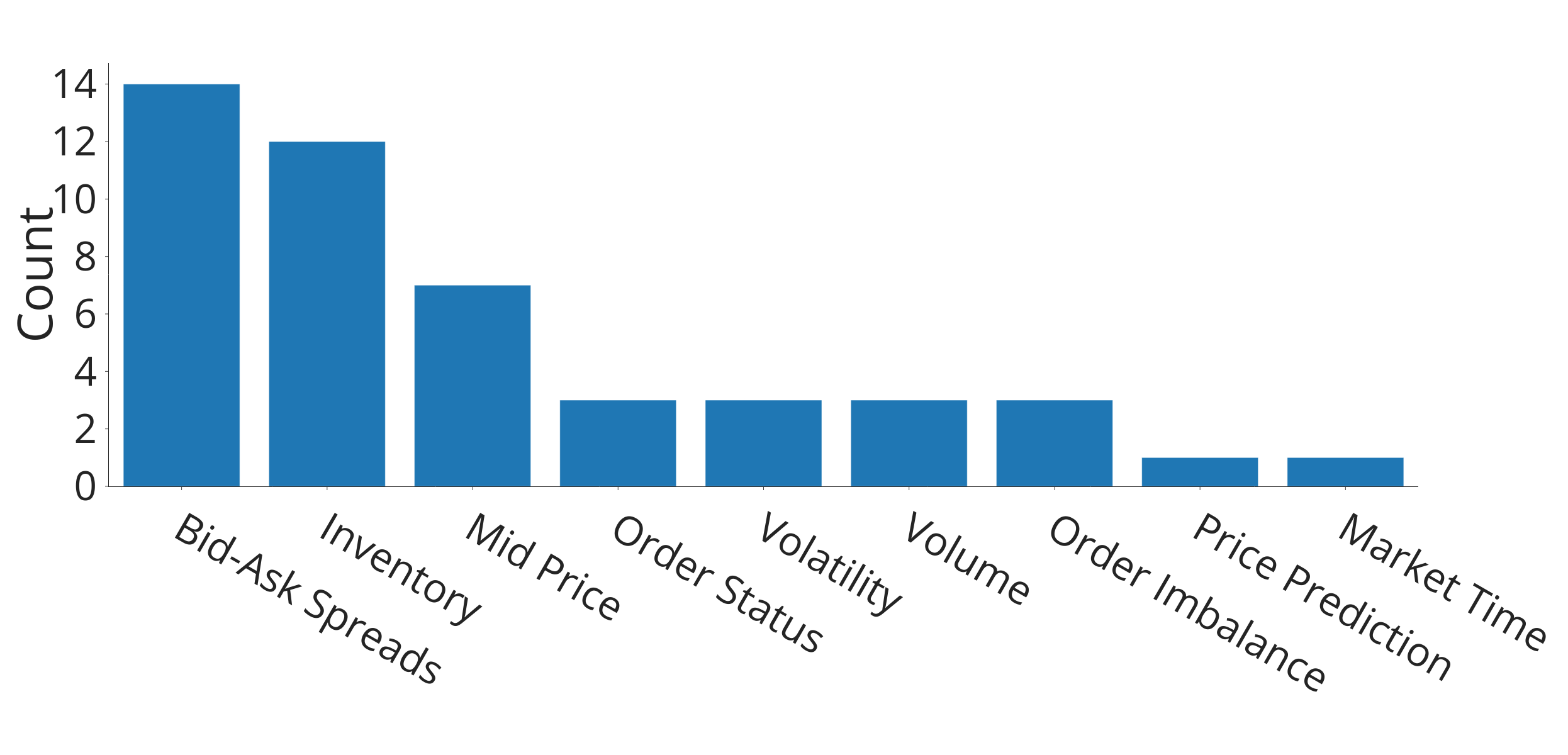}
        \caption{Tagged state space variables}
        \label{fig:state}
    \end{subfigure}
\end{figure}

Most analyzed references defined state spaces primarily based on market-level observations, as shown in \autoref{fig:state}.
specifically top-of-book quotes and N-depth book levels which align well with the real-time data available to market-making agents~\cite{He2023, Bakshaev2020}.
Additionally, agent inventory was also a common feature, reflecting the importance of managing risk and liquidity in market-making strategies~\cite{Patel2018, Ganesh2019}.
The tagged action spaces was mostly made up only one pair of quotes, with some references allowing the choice of multiple bid-ask levels.
Some references also used book levels as discrete action space variables, displayed in \autoref{fig:action}.

Finally, almost all reward structures, shown in \autoref{fig:reward},
were based on intraday or daily profit-and-loss scores, with either running liquidation, inventory penalties, or a combination of both.
The tagged action space variables also includes a separate ``Inventory'' variable that refers to portfolio strategies that output a target maximum inventory.
For implementations that use specific quantities per order, no additional tags were used,
as only 1 reference did not add quantities as part of the action space and instead used fixed quantities.

\subsection{Chosen Algorithms}
\label{subsec:chosen-algorithms}
Most references used Deep Q-Learning and Actor-Critic methods, leveraging neural networks for the optimal policy or state-value function estimation,
due to the high-dimensional nature of limit order books~\cite{Patel2018, Ganesh2019}.
The most used algorithms were PPO and DQN, as shown in \autoref{tab:table1},
both being the best performers in terms of convergence and generalization capabilities in the literature~\cite{Sun2022, Gasperov2021a}, with some references using DDQN variants.
The literature on model-free approaches overall demonstrated strong adaptability capabilities towards changing market conditions
while still maintaining acceptable training times.
Overall, the performed bibliographical review underscores the growing relevance and preference for reinforcement learning algorithms in the market-making domain,
particularly when combined with realistic spatial and temporal state representations.

\begin{table}
    \centering
    \begin{tabular}{|c|c|}
        \hline
        \textbf{Algorithm}           & \textbf{References} \\
        \hline
        Closed-Form Expression       & 8                   \\
        Deep (Double) Q-Learning     & 6                   \\
        Proximal Policy Optimization & 2                   \\
        Q-Learning                   & 2                   \\
        SARSA                        & 1                   \\
        Vanilla Policy Gradient      & 1                   \\
        SAC                          & 1                   \\
        \hline
    \end{tabular}
    \caption{Algorithms used in the references.}
    \label{tab:table1}
\end{table}

    \section{Methodology}
\label{sec:methodology}

The market-making problem addressed in this work involves designing an optimal trading policy for an agent using reinforcement learning.
The policy should be able to balance profit and risk, particularly under a market impact cost model,
with non-stationary dynamics and a complex state transition function.
The environment dynamics are modeled by an underlying limit order book (LOB) and
the details of the state transition operator and the process of obtaining the optimal decision policy to maximize cumulative rewards over time
are described in the following sections.

\subsection{A Formal Description of the Environment}
\label{subsec:formal-description-of-the-rl-environment}
In modeling the environment, we initially utilize a continuous-time, continuous-state Markov Chain,
and later transition to a discrete implementation to address computational space constraints.
Specifically, we model the environment as a Markov Decision Process (MDP),
defined as a 4-tuple $ (\mathcal{S}, \mathcal{A}, \mathbb{P}, R) $, where:

\begin{itemize}
    \item $\mathcal{S}$ is the set of states called the state space.
    \item $\mathcal{A}$ is the set of actions called the action space.
    \item $\mathbb{P}: \mathcal{S} \times \mathcal{A} \times \mathcal{S} \to [0, 1]$ is the transition probability function for the MDP.
    \item $R: \mathcal{S} \times \mathcal{A} \times \mathcal{S}' \rightarrow \mathbb{R}$ is the reward function associated with each state transition.
\end{itemize}

Then, the agent's interaction with the environment is made by chosing an action $a$ from the action space $\mathcal{A}$ in response to an observed state $s \in \mathcal{S}$,
according to a policy $\pi (s, a)$.
The agent's goal is to maximize cumulative rewards over time, which are obtained from the reward function $R$.
The return $G_t = \int_{t}^{T} \gamma^{k-t} R_{k} dk$ is used to estimate the value of a state-action pair by summing the rewards obtained from time $t$ to the end of the episode
(usually, as well as in our case, discounted by a factor $\gamma$ to favor immediate rewards),
where $R_{t}$ is the observed reward at time $t$ and $T$ is the time of the episode's end.

\subsubsection{Chosen State Space}
We choose a state space that tries to best incorporate the historical events of the limit order book into a single observable state using
market features deemed relevant and chosen through our initial bibliography research,
as well as guided by the works of
Gasparov et al. (2021)~\cite{Gasperov2021} and Gueant et al. (2017)~\cite{Gueant2017} on reinforcement learning for market-making.
We used the agent's current inventory and a set of indicators for the currently observed market state:
the Relative Strength Index (RSI); order imbalance (OI); and micro price (MP).
Additionally, for a fixed number $D$ of LOB price levels the pair $(\delta^d, Q^d)$, where $\delta^d$ is the half-spread distance for the level $d \leq D$,
and $Q^d$ the amount of orders posted at that level is added to the state as a set of tuples, for both the ask and bid sides of the book.
The state space can therefore be represented by the following expression:
\begin{equation*}
    \begin{aligned}
        s_{t} \in \mathcal{S} = \big\{ &\text{RSI}_t, \text{OI}_t, MP_{t}, \text{Inventory}_t, \\
        & \text{MA}_{10, t}, \text{MA}_{15, t}, \text{MA}_{30, t}, \\
        & (\delta_t^{d, ask}, Q_t^{d, ask})_{d=1}^{D}, \\
        & (\delta_t^{d, bid}, Q_t^{d, bid})_{d=1}^{D} \big\}
    \end{aligned}\label{eq:equation}
\end{equation*}
where $0 < t < T$, and $\text{MA}_{n, t}$ is the moving average of the price returns over the last $n$ time steps.
The indicators for our chosen market simulation framework are defined individually by values directly obtained from the observed LOB,
and serve as market state summaries for the agent to use:

\begin{itemize}
    \item \textbf{Order Imbalance (OI):} Measures the relative difference between buy and sell orders at a given time.
    \item \textbf{Relative Strength Index (RSI):} Momentum indicator that compares the magnitude of recent gains to recent losses to evaluate overbought or oversold conditions.
    \item \textbf{Micro Price (\( P_{\text{micro}} \)):} Weighted average of the best bid and ask prices, weighted by their respective quantities.
\end{itemize}
\begin{equation}
    \begin{aligned}
        \text{OI}_t &= \frac{Q_t^{\text{bid}} - Q_t^{\text{ask}}}{Q_t^{\text{bid}} + Q_t^{\text{ask}}}\\
        \text{RSI}_t &= 100 - \frac{100}{1 + \frac{\text{Average Gain}}{\text{Average Loss}}}\\
        \text{P}_{\text{micro}, t} &= \frac{P_t^{\text{ask}} Q_t^{\text{bid}} + P_t^{\text{bid}} Q_t^{\text{ask}}}{Q_t^{\text{bid}} + Q_t^{\text{ask}}}
    \end{aligned}
    \label{eq:features}
\end{equation}

where \( P_t^{\text{ask}} \) and \( P_t^{\text{bid}} \) represent the best ask and bid prices at time \( t \),
and \( Q_t^{\text{bid}} \) and \( Q_t^{\text{ask}} \) represent the total bid and ask quantities, respectively.
The \textit{Average Gain} and \textit{Average Loss} are computed over a rolling window (in our case, fixed 5 minute observation intervals),
and are defined as the average price increases and decreases during that window, respectively.
\( \text{OI}_t \in [-1, 1] \), with \( \text{OI}_t = 1 \) indicating complete dominance of bid orders, and \( \text{OI}_t = -1 \) indicating ask order dominance.
The RSI is bounded between 0 and 100, with values above 70 indicating overbought conditions and values below 30 indicating oversold conditions.
Overall, the chosen state space designed to capture the most relevant market features for the agent to make informed decisions is
not a novel approach per-se, containing common indicators used in market-making literature~\cite{Gueant2022, Selser2021a, FalcesMarin2022}.

\subsubsection{Chosen Action Space}

The control, or agent, interacts with the environment choosing actions from the set of possible actions,
such that $a \in \mathcal{A}$ in response to observed states $s \in \mathcal{S}$ according to a policy $\pi (s, a)$ which we will define shortly,
and the end goal is to maximize cumulative rewards over time.
The agent's chosen action impacts the evolution of the system's dynamics by inserting orders into the LOB that might move the observed midprice,
to introduce features of market impact into our model.

The action space $\mathcal{A}$ includes the decisions made by the agent at time $t$, specifically the desired bid and ask spreads pair
$\delta^{\text{ask}}, \delta^{\text{bid}}$ and the corresponding posted order quantities $Q^{\text{ask}}, Q^{\text{bid}}$:
$$
\mathcal{A} = \left\{ (\delta^{\text{ask}}, \delta^{\text{bid}}, Q^{\text{ask}}, Q^{\text{bid}}), \forall \delta \in \mathbb{R}^+, \forall Q \in \mathbb{Z}\} \right.
$$

\subsubsection{Episodic Reward Function and Returns}

The episode reward function $R_t \in \mathbb{R}$ reflects the agent's profit and inventory risk obtained during a specific time in the episode.
It depends on the spread and executed quantities, as well as the inventory cost and was choosen according to the best performing reward structures taken from the literature review.
For our model the utility chosen is based on a risk-aversion enforced utility function, specifically the \textit{constant absolute risk aversion (CARA)}~\cite{Arrow1965, Pratt1964}
with the virtual running Profit and Loss (PnL) as input.
It depends on the realized spread $\delta$ and the realized quantity $q$\footnote{Differs from the agent's posted order quantity $Q$, as $q$ is a stochastic variable dependent on the underlying market dynamics.},
and is computed as follows, where the penalty for holding large inventory positions is discounted by a factor of \( \eta \) from the final score:
\[
\begin{aligned}
    \text{Running PnL}_t &= \delta_t^{\text{ask}} q_t^{\text{ask}} - \delta_t^{\text{bid}} q_t^{\text{bid}} + \text{I}_t \cdot \Delta M_t, \\
    \text{Penalty}_t &= \eta \left( \text{Inventory}_t \cdot \Delta M_t \right)^+,\\
    \text{PnL}_t &\coloneqq \text{Running PnL}_t - \text{Penalty}_t
\end{aligned}
\]
Finally, the reward function is defined as the negative of the exponential of the running PnL, a common choice for risk-averse utility functions according to literature~\cite{Gueant2022, Selser2021a, FalcesMarin2022}.
The CARA utility function is defined as follows, where \( \gamma \) is the risk aversion parameter:
\[
    R_t = U(\text{PnL}_t) = -e^{-\gamma \cdot \text{PnL}_t},
\]

\subsection{State Transition Distribution}
\label{subsec:state-transition-distribution}

The previously mentioned transition probability density $P$ is given by a Stochastic Differential Equation expressed by the Kolmogorov
forward equation for Markov Decission Processes:

\begin{equation}
    \label{eq:equation2}
    \frac{\partial P(s', t | s, a)}{\partial t}  = \int_{\mathcal{S}} \mathcal{L}(x | s, a, t) P(s'| x, a, t) dx
\end{equation}

for all $s, s' \in \mathcal{S}$ and all times $t$ before market close $T$, that is, $t \le T$,
where $a$ is choosen by our control agent according to a policy $\pi (s)$.
$\mathcal{L}$ is the generator operator and governs the dynamics of the state transitions given the current time.

In continuous-time and state MDPs, the state dynamics is reflected by $\mathcal{L}$, and modern approaches to optimal control
solve analytically by obtaining a closed-form expression for the model's evolution equations, as originally tackled by
Avellaneda et al. (2008)~\cite{Avellaneda2008} and Gueant et al. (2017)~\cite{Gueant2017}.
or numerically by approximating its transition probabilities~\cite{Gueant2022, Selser2021a, FalcesMarin2022}.
Closed-form expressions for $\mathcal{L}$ are obtainable for simple environment models, by usually disconsidering market impact,
which is not the case for our proposed model, and solving for the generator operator is therefore outside the scope of this paper.
We describe in~\autoref{sec:implementation-and-model-description} our chosen approach to model the state transition distribution,
using the Proximal Policy Optimization (PPO) algorithm to approximate the optimal policy for the agent.

\subsection{Market Model Description and Environment Dynamics}
\label{subsec:market-model-description-and-environment-dynamics}
Our approach to the problem of market making leverages \textbf{online reinforcement learning} by means of a simulator that models the dynamics of
a limit order book (LOB) according to a set of stylized facts observed in real markets.
For our model of the limit order book the timing of events follows a \textit{Hawkes process}
to represent a continuous-time MDP that captures the observed stylized fact of clustered order arrival times.
The Hawkes process is a \textit{self-exciting process}, where the intensity \( \lambda(t) \) depends on past events.
Formally, the intensity \( \lambda(t) \) evolves according to the following equation:
\begin{equation*}
    \begin{aligned}
        \lambda(t) = \mu + \sum_{t_i < t} \phi(t - t_i),\\
        \phi(t - t_i) = \alpha e^{-\beta (t - t_i)},
    \end{aligned}
\end{equation*}

where \( \mu > 0 \) is the baseline intensity, and \( \phi(t - t_i) \) is the \textit{kernel function} that governs the decay of influence from past events \( t_i \).
A common choice for \( \phi \) is the exponential decay function, where \(\alpha\) controls the magnitude of the self-excitation and \(\beta\) controls the rate of decay.

The bid and ask prices for each new order are modeled by two separate \textit{Geometric Brownian Motion} processes
to capture the normally distributed returns observed in real markets.
The underlying partial differential equation governing the ask and bid prices are given by:
\begin{equation*}
    \begin{aligned}
        dX_{t}^{ask} &= (\mu_t + s_t) X_{t}^{mid} dt + \sigma dW_t,\\
        dX_{t}^{bid} &= (\mu_t - s_t) X_{t}^{mid} dt + \sigma dW_t,
    \end{aligned}
    \label{eq:gbm}
\end{equation*}
where $\mu$ is the price drift, $s_t$ is the mean spread, and $\sigma$ the price volatility.
The drift rate process follows a mean-reverting \textit{Ornstein-Uhlenbeck} process~\cite{Ornstein1930},
while the spread rate similarly follows a \textit{Cox-Ingersoll-Ross} process~\cite{Cox1985}, which, given that the
Feller condition $2\kappa_s \sigma_s^2 > 1$ is satisfied, ensures the spread remains positive:

\begin{equation*}
    \begin{aligned}
        d\mu_t &= \kappa (\mu - \mu_t) dt + \eta dW_t,\\
        ds_t &= \kappa_s (s - s_t ) dt + \sigma_s \sqrt{s_t} dW_t,
    \end{aligned}
    \label{eq:ou}
\end{equation*}

These two processes serve to model a market where return and spread regimes can vary over time, forcing the agent
to choose prices accordingly.
Whenever a new limit order that narrows the bid-ask spread or a market order arrive the midprice is updated to reflect the top-of-book orders.
The midprice $X_{t}^{mid}$ is therefore obtained by averaging the current top-of-book bid and ask prices:
\[
    X_{t}^{mid} = \frac{X_{t}^{ask} + X_{t}^{bid}}{2}
\]
and while there are no orders on both sides of the book, the midprice is either the last traded price,
observed or an initial value set for the simulation, in the given order of priority.
By analyzing the midprice process, we can make sure the returns are normally distributed, with mean $\mu_t$ and volatility $\frac{\sigma^2}{2}$,
assuring the model reflects a stylized fact of LOBs commonly observed in markets~\cite{Gueant2022}.

To model the underlying price volatility, we use a simple \textit{GARCH(1,1)} process, where the volatility process evolves according to \autoref{eq:garch}.
This choice is motivated by the fact that the GARCH(1,1) process is able to capture the clustering of volatility observed in real markets,
and can be easily calibrated to match the observed volatility of the desired underlying asset to be modeled.
\begin{equation}
    \begin{aligned}
        \sigma_t^2 = \omega + \alpha \epsilon_{t-1}^2 + \beta \sigma_{t-1}^2,
    \end{aligned}
    \label{eq:garch}
\end{equation}

Finally, the order quantities  $q_t^{\text{ask}}, q_t^{\text{bid}} \sim \text{Poisson}(\lambda_q)$
are modeled as Poisson random variables, where the arrival rate $\lambda_q$ is a constant parameter.
Our simulator was implemented using a Red-black tree structure for the limit order book,
while new orders follow the event dynamics described by the stochastic processes above.
The individual market regime variables, specifically the spread, order arrival density and return drift,
are sampled from the aforementioned set of pre-defined distributions, and inserted into the simulator at each event time step.

\subsection{Decision Process}
\label{subsec:decision-process}

In reinforcement learning, the Bellman equation is a fundamental recursive relationship that expresses the value of a
state in terms of the immediate reward and the expected value of subsequent states.
For a given policy $\pi$, the Bellman equation for the value function $V(s)$ with respect to the chosen reward function is:
\begin{equation*}
    \begin{aligned}
        V^\pi(s) &= \sum_{a \in \mathcal{A}} \pi(a \mid s) \sum_{s' \in \mathcal{S}} P(s'_{t+1} \mid s_{t}, a) \left[ R_t + \gamma V^\pi(s'_{t+1}) \right]\\
        V^\pi(s) &= \mathbb{E}_\pi \left[ R_t + \gamma V^\pi(s_{t+1}) \mid s_t = s \right]\\
    \end{aligned}
\end{equation*}

where $V(s)$ is the value function, representing the expected return (cumulative discounted rewards) starting from state $s$,
going to state $s'$ by taking action $a$ and continuing the episode, $R_t$ is the reward obtained from the action taken,
and $\gamma$ is the discount factor, used to weigh the value favorably towards immediate rather than future rewards.
The Bellman equation underpins the process of optimal policy derivation, where the goal is to find the optimal policy $\pi^*$,
that is, the control for our action space $\mathcal{A}$ that maximizes the expected return,
and can solved by maximizing for the value function:
\begin{equation*}
    \begin{aligned}
        V^*(s) &= \max_a \mathbb{E} \left[ R_t + \gamma V^*(s_{t+1}) \mid s_t = s, a_t = a \right]\\
        \pi^*(s) &= \arg \max_{a \in \mathcal{A}} \mathbb{E} \left[ R_t + \gamma V^*(s_{t+1}) \mid s_t = s, a_t = a \right]\\
    \end{aligned}
\end{equation*}
The Bellman equation is used to recursively define state values in terms of immediate rewards and expected value of subsequent states,
and for simpler environments, the equation can be solved directly by dynamic programming methods.
Such methods, like value iteration or policy iteration, although effective, become computationally expensive and sometimes unfeasible for large state spaces,
or require a model of the environment, which is not always available in practice.
Neural networks have been successfully used to approximate both the value function or the optimal actions directly,
demonstrating state-of-the-art results in various domains, including games, robotics, and finance
~\cite{He2023, Bakshaev2020, Patel2018, Ganesh2019, Sun2022, Gasperov2021a}.
Our simulation environment is complex enough that the state space is too large to store all possible state-action pairs,
and since we assume no knowledge of the underlying dynamics of the environment when training the agent, a model-free approach is necessary.

\subsubsection{Generalized Policy Iteration and Policy Gradient}
\label{subsubsec:gpi}
Generalized Policy Iteration algorithms are a family of algorithms that combines policy evaluation and policy improvement steps iteratively,
so that the policy is updated based on the value function, and the value function is updated based on the policy.
While simpler methods directly estimate state values or state-action values, such as Q-learning, more complex methods
such as Actor-Critic algorithms, estimate both the policy and the value function.
The critic estimates the value function $V^\pi$, or $A^{\pi} = Q^{\pi} - V^{\pi}$, if using the advantage function, while the actor learns the policy $\pi$.
We use a \textbf{policy gradient} approach, where the agent optimizes the policy directly by maximizing the expected return,
through consecutive gradient ascent steps on the policy parameters given episodes of experience.
For policy gradient methods, the following equations express the gradient ascent step for the
policy parameters $\theta$ and the value function parameters $\phi$, using a $\alpha$ learning rate and the gradient operator $\nabla$:
\begin{gather*}
    \nabla_{\theta} J(\theta) = \mathbb{E}_{\tau \sim \pi_{\theta}} \left[ \sum_{t=0}^{T} \nabla_{\theta} \log \pi_{\theta}(a_t \mid s_t) A^{\pi}(s_t, a_t) \right]\\
    \nabla_{\phi} J(\phi) = \mathbb{E}_{\tau \sim \pi_{\theta}} \left[ \sum_{t=0}^{T} \nabla_{\phi} \left( V^{\pi}(s_t) - R_t \right)^2 \right]\\
    \pi_{\theta} \leftarrow \pi_{\theta} + \alpha \nabla_{\theta} J(\theta), \phi \leftarrow \phi + \alpha \nabla_{\phi} J(\phi)
\end{gather*}
\begin{figure}
    \centering
    \includegraphics[width=1\columnwidth]{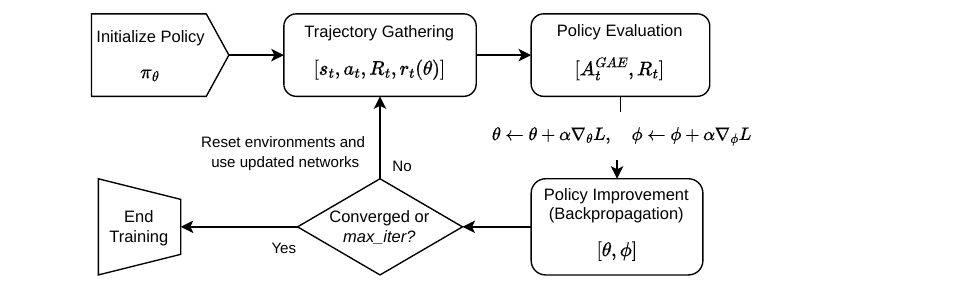}
    \caption{Diagram of the Generalized Policy Iteration loop for the implemented agent.}
    \label{fig:gpi}
\end{figure}

In the context of \textbf{Proximal Policy Optimization (PPO)}
the model is trained to improve its policy and value function by maximizing the expected return,
according to a clipped surrogate objective function, which ensures that the policy does not deviate excessively from the previous policy,
promoting stability during training.
For \textbf{PPO}, the loss function used for the policy gradient ascent is defined as:
\begin{gather*}
    L^{\text{actor}}(\theta) = \mathbb{E}_t \left[ \min \left( r_t(\theta) \hat{A}_t, \text{clip}(r_t(\theta), 1 - \epsilon, 1 + \epsilon) \hat{A}_t \right) \right],\\
    \hat{A}_t = \sum_{l=0}^{\infty} (\gamma \lambda)^l \delta_{t+l}.
\end{gather*}
where \( r_t(\theta) \) is the probability ratio between the new and old policies and $\hat{A}_t$ is the advantage function.
The critic loss is the mean square error between the predicted value $V(s_t)$ and the actual return $R_t$, defined as:
\begin{gather*}
    L^{\text{critic}} = \mathbb{E}_t \left[ \left( V(s_t) - R_t \right)^2 \right],\\
\end{gather*}
Algorithmically, this is implemented by gathering trajectories from the environment and estimating the expression above using finite differences
as shown in \hyperref[alg:ppo]{Algorithm~\ref{alg:ppo}}.
\begin{algorithm}
    \begin{algorithmic}[1]
        \Require Policy $\pi_\theta$, value function $V_\phi$, trajectories $\tau$, epochs $K$, batch size $B$
        \State \textbf{Compute Advantage Estimation:}
        \State $\hat{A}_t = \sum_{l=0}^{\infty} (\gamma \lambda)^l \delta_{t+l}$
        \State \textbf{Compute Returns:} $R_t = \hat{A}_t + V(s_t)$
        \State \textbf{Construct dataset} $\mathcal{D} = (s_t, a_t, \hat{A}_t, R_t, \log \pi_\theta(a_t | s_t))$
        \For{epoch $k = 1$ to $K$}
            \For{minibatch $(s, a, \hat{A}, R, \log \pi_{\theta_{\text{old}}}(a | s))$ in $\mathcal{D}$}
                \State \textbf{Policy Update:}
                \State Compute probability ratio:
                \(
                r_t(\theta) = \frac{\pi_\theta(a | s)}{\pi_{\theta_{\text{old}}}(a | s)}
                \)
                \State Compute loss separately for actor and critic:
                \begin{equation*}
                    \begin{aligned}
                        L^{\text{actor}} &= \mathbb{E} \left[ \min(r_t(\theta) \hat{A}_t, \text{clip}(r_t(\theta), 1 - \epsilon, 1 + \epsilon) \hat{A}_t) \right]\\
                        L^{\text{critic}} &= \mathbb{E} \left[ (R_t - V(s_t))^2 \right]\\
                    \end{aligned}
                \end{equation*}
                \State Perform backpropagation and gradient update:
                \[
                    \theta \gets \theta + \alpha \nabla_\theta L^{\text{actor}}, \quad \phi \gets \phi + \alpha \nabla_\phi L^{\text{critic}}
                \]
            \EndFor
        \EndFor
    \end{algorithmic}
    \caption{Actor-Critic with PPO Updates}
    \label{alg:ppo}
\end{algorithm}

Each pass updates both the policy parameters $\theta$ and the value function parameters $\phi$
by applying the usual backward propagation pass for neural networks to minimize the loss function according to a chosen optimizer.
A more detailed explanation of the PPO algorithm can be found in the original paper by Schulman et al. (2017)~\cite{Schulman2017}.
Our implementation of the algorithm is a distributed-parallel version of the original version,
where multiple learners interact with the environment in parallel and share the experience to update the policy and value function.
Short of the V-trace algorithm, this approach is similar to the IMPALA algorithm~\cite{Espeholt2018}.
Besides the proposed approach for simulating dynamic limit order book environments,
we also implement a hybrid neural network for the agent's policy and value functions,
composed of a sequence of self-attention layers to capture the spatial dependencies between the different levels of the LOB,
concatenated with dense layers to process the market features, as discussed in~\autoref{sec:implementation-and-model-description}.

\subsubsection{Benchmark Closed-Form Expression for Simplified Model}
To ensure the agent's performance is able to capture the complexity of the environment, we use a benchmark closed-form expression
to compare the agent's performance against.
We chose the closed-form expression for the optimal bid-ask spread pair as proposed by Avellaneda et al. (2008)~\cite{Avellaneda2008},
which is given by the following expression:
\begin{equation}
    \begin{aligned}
        \delta^* &= \frac{\sigma}{\sqrt{2}} \text{erf}^{-1} \left( \frac{1}{2} \left( 1 + \frac{\mu}{\sigma} \right) \right),\\
        &\text{erf}(x) = \frac{2}{\sqrt{\pi}} \int_{0}^{x} e^{-t^2} dt.
    \end{aligned}
    \label{eq:avellaneda}
\end{equation}
where $\sigma$ is the volatility, $\mu$ is the mean spread, and $\text{erf}^{-1}$ is the inverse error function.
The error function $\text{erf}(x)$ serves simply as a measure of the spread of the normal distribution.

The closed-form expression is derived from a simple market model which assumes normally distributed non-mean reverting spreads,
constant executed order size and exponentially distributed event time dynamics.
As it depends on a simpler model for the market to be implemented, our expectations are that the agent will under-perform in a more complex environment.
The optimal bid-ask spread pair according to \autoref{eq:avellaneda} is then given by $p_\text{bid} = \mu - \delta^*$ and $p_\text{ask} = \mu + \delta^*$.
A fixed quantity of $1$ quoted share as originally proposed is used.

    \section{Implementation and Model Description}
\label{sec:implementation-and-model-description}

As previously stated in \autoref{sec:methodology}, the chosen model follows the Actor-Critic architecture,
with the Actor network learning the policy and the Critic network learning the value function.
For the actor network input, we separate the state space into two tensors: the market features and the LOB data.
The market features contain general high-level information, including the microprice, 10, 15 and 30-period moving averages,
agent inventory, and the Relative Strength Index (RSI) and Order Imbalance (OI) indicators.
The LOB data contains the ordered $N$th best bid-ask price and volume pairs.

\begin{figure}
    \centering
    \includegraphics[width=.9\columnwidth]{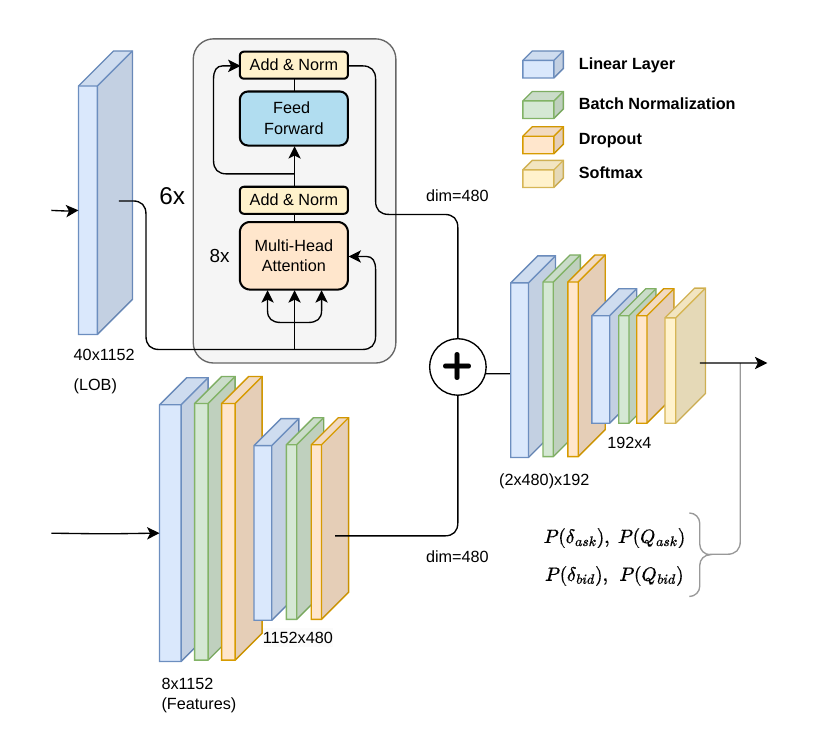}
    \caption{Actor Network Architecture}
    \label{fig:actor-architecture}
\end{figure}

We use a sequence of self-attention layers to capture the spatial dependencies between the different levels of the LOB,
while the market features are passed through a dense layer and concatenated with the output of the self-attention layers,
before being passed through the final dense layers as shown in \autoref{fig:actor-architecture}.
We implement the Critic network as a simple feed-forward neural network with two hidden layers, of 128 and 64 units, respectively,
with the same input tensors as the Actor network.

To train our policy and value networks, we use the aforementioned PPO algorithm\cite{Schulman2017},
which is a model-free, on-policy approach to optimize the policy directly, as discussed in~\hyperref[subsubsec:gpi]{Subsection~\ref{subsubsec:gpi}}.
Our training loop shown in \hyperref[alg:algorithm]{Algorithm~\ref{alg:algorithm}} consists of collecting trajectories,
computing the Generalized Advantage Estimation (GAE), which is used instead of the usual returns $G_t$.
The chosen approach can be classified as \textbf{online reinforcement learning}, where the agent learns from interactions with the environment,
without the need for a pre-existing dataset, as opposed to offline reinforcement learning, where the agent learns from a static dataset.
Additionally, we use a replay buffer to store the trajectories and sample mini-batches for training the policy and value networks.
Since the replay buffer is discarded immediately after a small fixed number of epochs, the approach is closer to \textbf{on-policy learning},
as the data collected by the policy is the same as the data used for gradient updates.

\begin{algorithm}
    \begin{algorithmic}[1]
        \Require Environment, PPO model, optimizer, number of episodes $num\_episodes$
        \For{each episode in range $num\_episodes$}
            \State Reset environment and observe initial state $s$
            \For{each timestep until episode ends}
                \State Select action $a \sim \pi_{\theta}(s)$
                \State Observe reward $R_t$ and next state $s'$
                \State Store transition $(s, a, r)$ in the trajectory buffer
                \State Set $s \leftarrow s'$
            \EndFor
            \State \textbf{Compute GAE and Returns} // Policy evaluation
            \State \textbf{Update parameters} $\boldsymbol{\theta}$ \textbf{and} $\boldsymbol{\phi}$ // Policy improvement
        \EndFor
    \end{algorithmic}
    \caption{Training Loop}
    \label{alg:algorithm}
\end{algorithm}

    \section{Experiments and Results}
\label{sec:realized-experiments-and-results}

\subsection{Experimental Setup}
\label{subsec:experiment-setup}

We trained the agent for a total of 10,000 episodes, with an average of 390 observations per episode --- or 1 event per market minute.
For training, we used a single NVIDIA GeForce RTX 3090 GPU with 24GB of memory, and an AMD Ryzen 9 5950X CPU with 16 cores and 32 threads,
and 32GB of DDR4 RAM.
We used the Adam optimizer with a learning rate of $3 \times 10^{-4}$ for both the policy and value networks, with 64 epochs per trajectory.
The discount factor $\gamma$ was set to 0.9, the GAE parameter $\lambda$ was set to 0.85, and the PPO clipping parameter $\epsilon$ was set to 0.25.
The entropy coefficient was set to $1.2\times10^{-3}$, and the batch size set to 256 samples per episode/update.
The hyperparameters were obtained through a grid search optimization process.

For our market model, we chose the following parameters for each process, generating 390 samples before each episode to heat up the book:
\begin{itemize}
    \item The order arrival rate was set to $\lambda = 1$, with clustering parameters $\alpha = 0.1$ and $\beta = 0.1$ (yearly volume of about $100.000$ orders).
    \item Mean spread was set to $s = 0.1$ (10 market price ticks, in our case, 10 cents) and annualized price drift to $\mu = -0.02$ ($-2\%$), respectively.
    \item The price volatility parameters were set to $\omega = 0.5$, $\alpha = 0.1$, and $\beta = 0.1$,
    where $\omega$ is the constant term, $\alpha$ the autoregressive term, and $\beta$ the moving average term.
    \item The initial midprice was arbitrarily set to $100$.
\end{itemize}

\subsection{Results Obtained}
\label{subsec:experiment-results}

To evaluate the financial performance of the trained reinforcement learning agent, we analyzed the agent's
financial return, return volatility, and the Sortino ratio.
The results were averaged over $10^2$ trajectories using the same hyperparameters used for training.

\begin{figure}[t]
    \centering
    \begin{minipage}{\columnwidth}
        \centering
        \includegraphics[width=1\textwidth]{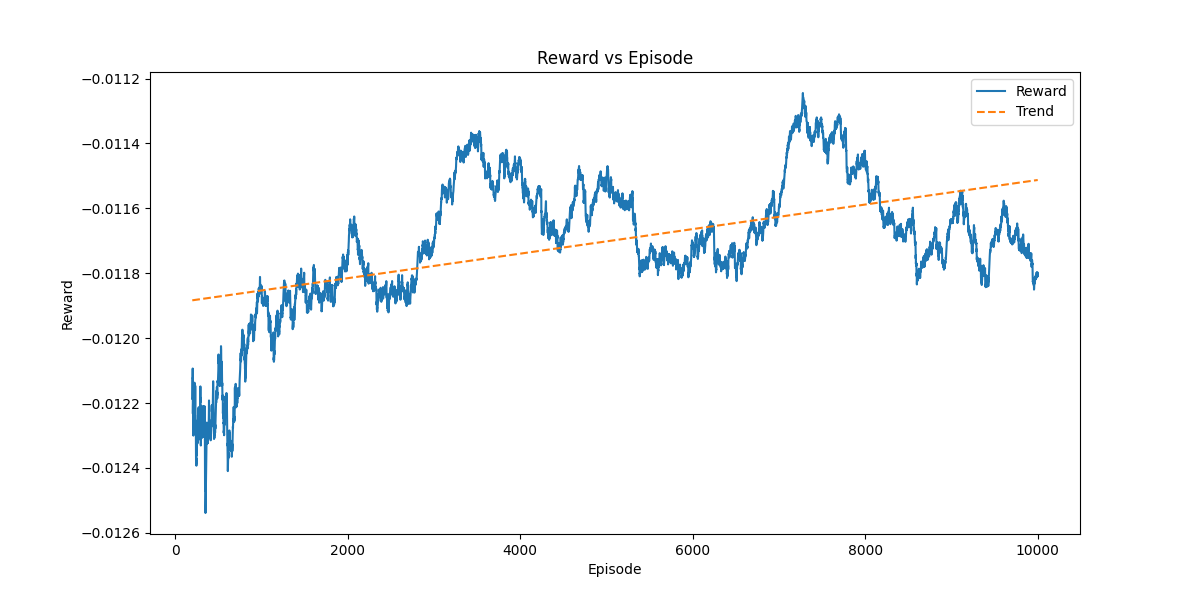}
        \caption{Exponential moving average of the training reward per episode, with a linear trend line.}
        \label{fig:average-reward-moving-average}
    \end{minipage}
    \vspace{0.04\textwidth} 
    \begin{minipage}{\columnwidth}
        \centering
        \includegraphics[width=1\textwidth]{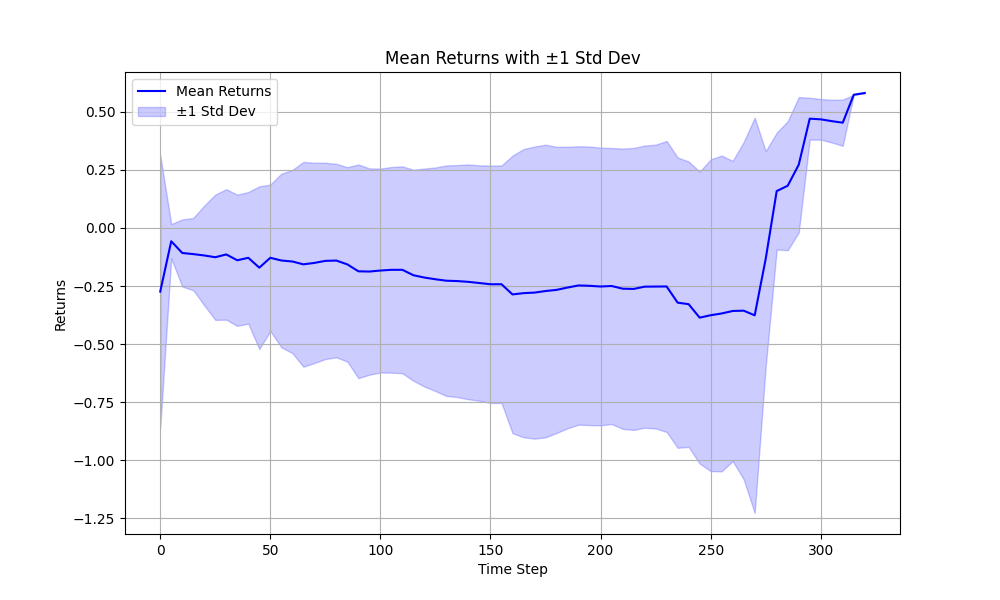}
        \caption{Financial return, averaged over 100 trajectories with a 1 standard deviation confidence interval.}
        \label{fig:average-financial-return}
    \end{minipage}
\end{figure}

As shown in Table~\ref{tab:test-results}, the reinforcement learning agent exhibited a mean financial return of $5.203 \times 10^{-5}$
(annualized return of about $+1.31\%$.), an almost neutral performance under adverse market conditions,
while the benchmark agent had a mean financial return of $3.037 \times 10^{-5}$ (annualized return of about $+0.76\%$),
underperforming under the same conditions, but closely matching the expected performance.
For the simple long-only strategy, the mean financial return was $-2.206 \times 10^{-5}$ (annualized return of about $-0.56\%$),
demonstrating how the agent's learned policies allowed the RL paradigm to outperform a simple long-only strategy under adverse market conditions.
The Sortino ratio for the RL-agent was $0.7497$, outperforming both benchmark agents.
Observing the reward curve during training, we can see a steady increase in the reward over time, as shown in Figure~\ref{fig:average-reward-moving-average},
indicating that the agent was learning to maximize its cumulative rewards over time, as expected.

The agent's processing time was also measured, as latency is a critical metric for real-world applications,
with the actor network taking on average $2.90$ms per episode, and the critic network taking $0.21 \times 10^{-4}$ seconds per episode,
well within the acceptable range for medium- and low-frequency trading.
Further training could be performed if necessary to improve the agent's performance, but the results obtained were already considered satisfactory
considering the fabricated adverse market conditions.

\begin{table*}
    \centering
    \small

    \begin{tabular}{|c|c|c|}
        \hline
        \textbf{Training}      & \textbf{Metric}                       \\
        \hline
        Training Time          & $\SI{112740000}{\milli\second}$       \\
        Time per Episode       & $845.8 \pm \SI{104.4}{\milli\second}$ \\
        Processing Time Actor  & $2.90 \pm \SI{1.0}{\milli\second}$    \\
        Processing Time Critic & $0.21 \pm \SI{0.03}{\milli\second}$   \\
        \hline
    \end{tabular}
    \caption{Test Results}
    \label{tab:test-results}
    \centering
    \vspace{0.5cm}
    \small

    \begin{tabular}{|c|c|c|c|}
        \hline
        \textbf{Test} & \textbf{RL-Agent}      & \textbf{Stoikov}       & \textbf{Long-Only}      \\
        \hline
        Mean Return   & $5.203 \times 10^{-5}$ & $3.038 \times 10^{-5}$ & $-2.207 \times 10^{-5}$ \\
        Volatility    & $1.178 \times 10^{-4}$ & $1.714 \times 10^{-4}$ & $2.396 \times 10^{-3}$  \\
        Sortino Ratio & 0.7497                 & 0.4271                 & -0.0079                 \\
        \hline
    \end{tabular}
    \caption{Training Results}
    \label{tab:training-results}
\end{table*}


Overall, the training process maintained an increasing reward curve and finished in about 3 hours and 8 minutes.
The financial return of the RL-agent, as shown in Figure~\ref{fig:average-financial-return}, was stable around zero,
and even though we expected a slightly negative return due to the highly volatile market conditions and
somewhat low drift, the positive but close to neutral performance was still considered a favorable outcome,
especially when compared to the benchmark agent's performance.
The calculated Sortino ratio of $0.7497$ further supports the agent's performance under adverse market conditions,
when compared to the benchmark agent scores of $0.4271$ and $-0.0079$ for the Stoikov and Long-Only strategies, respectively.

The agent demonstrated being capable of adapting to changing market dynamics and outperform simpler strategies,
where the proposed simulator provided a realistic environment for training under non-stationary market conditions.
We confirmed our initial hypothesis that stochastic dynamic environments can effectively be used to simulate
a market environment with changing regimes, and that reinforcement learning agents can learn to adapt to these conditions,
instead of using historical data with no market impact or sample-biased generative models for training.

    \section{Conclusion}
\label{sec:conclusion}
We have presented the design and implementation of a reinforcement learning agent aiming to show
the effects of adverse market conditions and non-stationary environments on control agents for market-making.
As discussed in~\autoref{subsec:market-model-description-and-environment-dynamics},
our approach for a environment models the dynamics of a limit order book (LOB)
according to a set of parameterizable stochastic processes configured to mimic observed stylized facts in real markets
The resulting market model replicates stylized facts for the midprice, spread, price volatility, and order arrival rate,
as well as the impact of market orders on the agent's inventory, return standard deviation and end-of-day Profit and Loss score.

The fine-controlled dynamics where the agent interacts with the environment allows us to model the effects of market impact and inventory risk on the agent's performance,
and use it to evaluate the agent's performance under adverse market conditions, providing a more realistic training environment for RL agents in market-making scenarios.
The proposed methodology was able to capture the effects of market impact and inventory risk on the trading agent's performance,
and affected the resulting agent's decision-making policies, as shown in~\autoref{subsec:experiment-results},
which we verified by comparing the agent's performance against simpler benchmark agents that
do not account for these effects in their decision-making policies.

Prospective extensions of the presented work and future research on existing market model simulators include
developing hybrid world models to combine both model-based and model-free approaches and leverage the capability of RL agents
to learn from non-stationary environments and historical observations of real markets.
Hybrid world models could further improve the capability of RL agents in adapting to changing market conditions
and could provide a more realistic training environment for market-making agents.

    \vspace{-1em}
    \section*{Acknowledgments}
    The author(s) gratefully acknowledge support from the São Paulo Research Foundation (FAPESP grant no. 2023/16028-3).

    \bibliographystyle{plain}
    \bibliography{references}

\end{document}